\documentclass[reprint,aps,prl,superscriptaddress]{revtex4-2}
\usepackage{amsmath}
\usepackage{amsfonts}
\usepackage{amssymb}
\usepackage{graphicx}
\usepackage{tabularx}
\usepackage{mathtools}
\usepackage{xcolor}
\usepackage{float}

\begin{document}
\title{Reference Lattice, Sound, Stiffness, and Magnetic Transitions of Ising Monolayers}

\author{John M. Davis}
\affiliation{Department of Physics and MonArk NSF Quantum Foundry, University of Arkansas, Fayetteville, AR 72701, USA}
\author{Amador Garc{\'\i}a-Fuente}
\affiliation{Departamento de F{\'\i}sica and Centro de Investigaci\'{o}n en Nanomateriales y Nanotecnolog\'{\i}a, Universidad de Oviedo–CSIC, 33940 El Entrego, Spain}
\author{Jaime Ferrer}
\affiliation{Departamento de F{\'\i}sica and Centro de Investigaci\'{o}n en Nanomateriales y Nanotecnolog\'{\i}a, Universidad de Oviedo–CSIC, 33940 El Entrego, Spain}
\author{Salvador Barraza-Lopez}
\affiliation{Department of Physics and MonArk NSF Quantum Foundry, University of Arkansas, Fayetteville, AR 72701, USA}

\begin{abstract}
A reference lattice, away from which elastic distortions induced by the spin texturing of 2D magnets take hold, is motivated from a picture of pairwise Biot-Savart interactions among identical solenoids that either elongate or compress a (``zero-current'') spring lattice. Applied to a paradigmatic CrSiTe$_3$ monolayer (ML), the reference is given by the average between the atomic positions of FM and N\'eel AFM lattices; such an atomic disposition permits understanding structural distortions and elastic energies due to magnetism readily. Furthermore, the anisotropic speed of sound in the magnetic ground state explains an observed anisotropy of vibrational frequencies on similar magnets. Elastic stiffness constants are reported, too. Magnetic energies in four Ising structural configurations were calculated, and the strain needed for those 2D magnets to undergo an AFM to FM quantum phase transition was determined as well.
\end{abstract}

\date{\today}
\maketitle

Magnetism is relevant in Condensed Matter Physics \cite{kittel2018introduction}, Statistical Mechanics \cite{Kivelson}, and Quantum Field Theory \cite{Fradkin}. Depending on whether spins align perpendicularly to a plane, parallel to it, or take an arbitrary orientation in three dimensions, magnetic materials are studied with {\em Ising} \cite{Ising}, {\em Potts} (also called {\em XY}) \cite{Potts}, and {\em Heisenberg} \cite{Heisenberg} models, respectively \cite{book}.

Most layered magnets \cite{McGuire,cit:burchN2018} are binary transition metal halides ({\em e.g.}, CrI$_3$, CrI$_2$, CrCl$_3$, Nb$_3$Cl$_8$, and ScCl), binary transition metal chalcogenides ({\em e.g.}, VSe$_2$, and Cr$_3$Se$_4$), binary transition metal compounds (Co$_2$P, CrB$_2$, MnO$_2$, and FeSi$_2$), binary materials with spin-unpaired $p$- or $f$-electrons (GdI$_2$, or K$_2$N), or ternary transition metal compounds (MPS$_3$, MPSe$_3$, CrSiTe$_3$, CrGeTe$_3$, Fe$_5$GeTe$_2$, MnBi$_2$Te$_4$, CrSBr, and CrSiRe$_3$) \cite{APLReview}. The Curie temperature ($T_C$) at which those phases turn paramagnetic (of the order of tens of meV per formula unit) is tunable by thickness \cite{genome}, doping \cite{doping1}, 
electric field \cite{deng}, 
and strain \cite{cit:zhangNL2021,cit:niNN2021}.

Experimental evidence for {\em magnetoelastic couplings} was first reported in bulk CrGeTe$_3$ \cite{cit:tian2DM2016}, and confirmed in FePS$_3$, FeSe$_2$ \cite{Crommie}, FePS$_3$, CoPS$_3$, and NiPS$_3$ \cite{cit:houmesNC2023}: lattice parameters change across the ferromagnetic (FM) to paramagnetic phase transition \cite{cit:vijaySR2023}. From a theoretical perspective, Ref.~\cite{cit:sivadasPRB2015} reported four Ising magnetic structures for CrSiTe$_3$ and CrGeTe$_3$ MLs, {\em all having a honeycomb lattice}. Recently, the structural anisotropy of this family of layered magnets was experimentally established \cite{cit:houmesNC2023} through the anisotropy of the fundamental vibration frequency of bulk MPS$_3$ (M=Fe, Co, or Ni).

The magnetic ground state of CrSiTe$_3$ is strongly dependent on electronic correlations and on the number of MLs. At the ML limit, DFT calculations based on LDA exchange-correlation yield a FM ground state \cite{cit:linJMCC2016}, while those utilizing a PBE exchange correlation yield a zigzag AFM magnetic ground state \cite{cit:sivadasPRB2015}. Experimentally, a FM ground state has been determined at the monolayer limit \cite{Jaime1,fujita}. Presenting a collection of methodologies to couple structural distortions to magnetism, and exemplified on CrSiTe$_3$ MLs, this Letter (i) introduces a suitable non-magnetic {(NM)} reference configuration for 2D magnets, (ii) it establishes the crystal symmetries (layer and point groups) of the four ML structures, (iii) presents their anisotropic sound velocities and out-of-plane vibrational modes, and (iv) contributes stiffness parameters (which require knowledge of the point group to be determined appropriately). (v) Knowledge of the proper reference lattice permits calculating elastic energies, which are all smaller than the energy barriers among magnetic phases. (vi) The free energy invites a discussion of magnetic quantum phase transitions.

The original motivation for this work was to demonstrate that two of the four magnetic phases reported by Sivadas {\em et al.}~\cite{cit:sivadasPRB2015} lacked hexagonal symmetry. For this reason, we used calculation parameters that closely matched theirs. Our calculations were performed with the SIESTA density functional theory (DFT) code \cite{cit:siesta1, cit:siesta2} using the PBE exchange-correlation functional \cite{cit:pbe} and a Hubbard correction \cite{cit:dudarev} $U_{\textrm{eff}}=4.0$ eV for $d$-electrons. Energy, force and stress tolerances were set to $10^{-6}$ eV, $10^{-6}$ eV/\AA{}, and $10^{-3}$ GPa, respectively. A $k$-point mesh of $30\times30\times1$ ($30\times18\times1$) was utilized for FM and N\'{e}el AFM (zigzag AFM and stripy AFM) unit cells (u.c.s); phonon dispersions were generated using a $5\times5\times1$ ($5\times3\times1$) supercell and a $6\times6\times1$ $k$-point mesh. The out-of-plane lattice vector was set to $\mathbf{c}=(0,0, 20)$ {\AA}. Lattice stiffness (``force constant'') tensors $\mathbf{K}_{ij}$ and elastic energies were computed within the frozen-phonon approximation \cite{cit:feynman,frozenphonon} (we used $\pm 2.5\times10^{-2}$ {\AA} atomic displacements along the Cartesian directions to calculate forces). Structural optimizations and phonon calculations were performed with spin-polarization. Stiffness tensors were diagonalized for the FM and N\'eel AFM structures. Our energetics and structures match those obtained using other DFT codes; see Supplemental Materials for details \cite{SI}.

Our code GROGU \cite{Jaime2}---that maps Kohn-Sham Hamiltonians {\em with noncollinear spin} to classical Heisenberg models---was used to find exchange and anisotropy tensors from the magnetic force theorem~\cite{Solovyev2021}. There, energy changes induced by small twists in the relative spin magnetization are related to exchange tensors {\em via} perturbation theory \cite{JaimeMethod1,Jaime2,Jaime3}. The Heisenberg model:
\begin{equation}\label{eq:heisenberg}
    H_M = \sum_{i<j}{\left\{\mathbf{e}_i \mathbf{J}_{ij}\mathbf{e}_j + \mathbf{D}_{ij}\cdot\left(\mathbf{e}_i\times\mathbf{e}_j\right)\right\}}
        + \sum_{i}\mathbf{e}_i\mathbf{A}_i\mathbf{e}_i
\end{equation}
contains symmetric ($\mathbf{J}_{ij}$) and Dzyaloshinskii-Moriya ($\mathbf{D}_{ij}$) \cite{Dzyaloshinsky1958,Moriya1960} exchange interactions between sites $i$ and $j$, as well as single-ion magnetic anisotropy ($\mathbf{A}_i$). Spin momentum vectors ($\mathbf{S}_i$) were normalized to 1 ($\mathbf{e}_i$).

\begin{figure}[tb]
    \centering
    \includegraphics[width=0.45\textwidth]{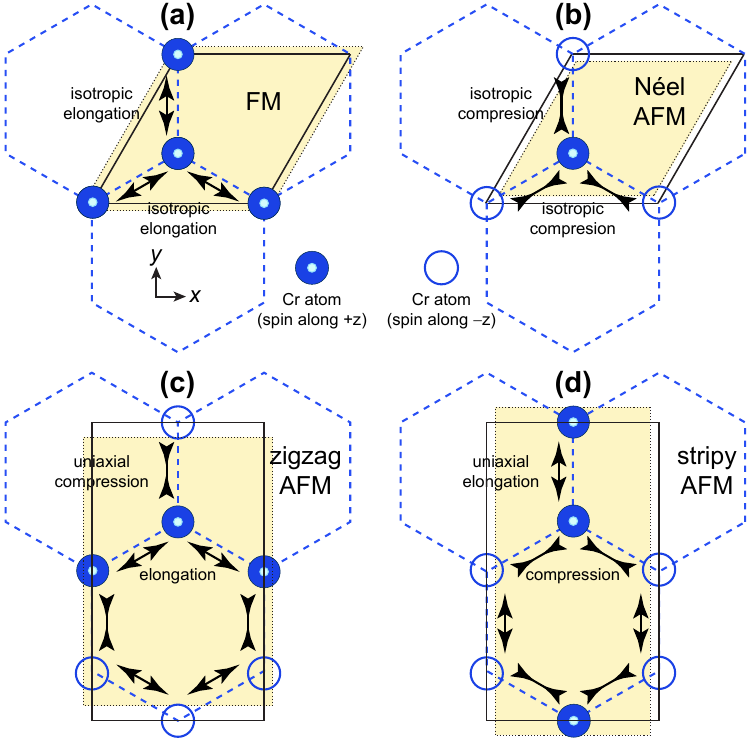}
    \caption{Magnetic forces on a 2D ``solenoid lattice'' under Ising magnetism. (a) Parallel moments repel isotropically. (b) Anti-parallel moments attract isotropically. Other Ising textures create: (c) elongation (compression) along $x$ ($y$), or (d) compression (elongation) along the $x$ ($y$) direction.
\label{fig:F0}}
\end{figure}

\begin{figure}[tb]
    \centering
    \includegraphics[width=0.45\textwidth]{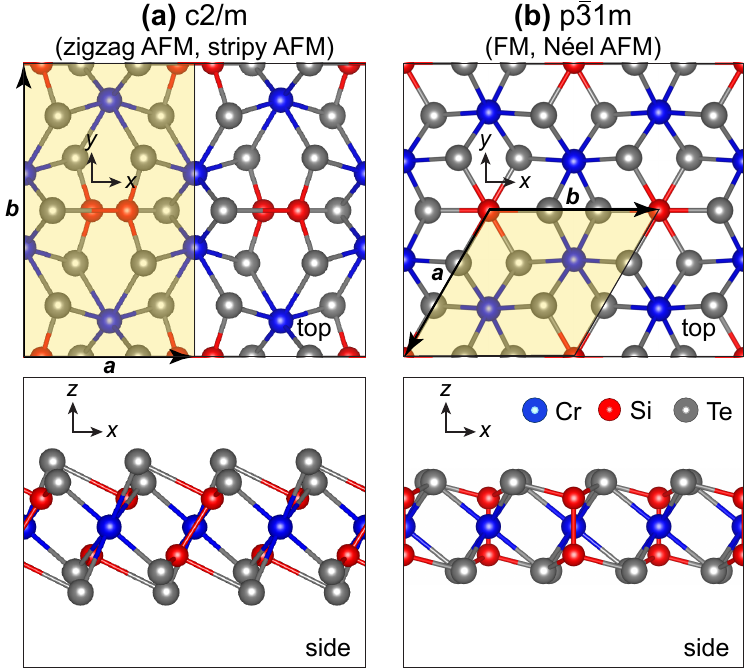}
    \caption{(a) Anisotropically distorted and (b) isotropically distorted CrSiTe$_3$ MLs. The distortions in (a) were magnified to better understand reduced symmetries. See Ref.~\cite{SI}.\label{fig:F1}}
\end{figure}

To a first approximation, collinear {\em pairwise} magnetic interactions between atoms $i$ and $j$ on an Ising 2D lattice are described by this Hamiltonian: $J(r_{ij})\,S_{i,z}S_{j,z}$, where $J(r_{ij})$ is the exchange interaction between atoms $i$ and $j$, separated by vector $\mathbf{r}_{ij}$ with $r_{ij}=|\mathbf{r}_{ij}|$. The dependence of $J(r_{ij})$ with the distance $r_{ij}$ between atoms $i$ and $j$ creates a magnetic force:
\begin{equation}\label{eq:1}
{\bf F}_{\mathbf{r}_{ij}}=-{\bf \nabla}_{\mathbf{r}_{ij}} J(r_{ij})\,S_{i,z}\,S_{j,z}
\end{equation}
of alternating sign for parallel and anti-parallel spin orientations, leading to opposite {\em magnetostrictive} effects. Assuming spins are independent of the distance between magnetic atoms, and for spins oriented ferromagnetically, a decrease of $J(r_{ij})$ as the separation among two atoms increases leads to an isotropic repulsive force of the parallel magnetic ions [Fig.~\ref{fig:F0}(a)]. Likewise, when spins are antiparallel, a decrease of $J(r_{ij})$ as the separation $r_{ij}$ increases induces an isotropic attraction [Fig.~\ref{fig:F0}(b)].

Arrangements of Ising magnetic moments with lower symmetry distort the lattice {\em anisotropically} \cite{cit:houmesNC2023}: The magnetic texture shown in Fig.~\ref{fig:F0}(c) has opposite Ising orientations along vertically-separated zigzag lines, for a ``zigzag AFM'' configuration. Since parallel magnetic moments repel and antiparallel ones attract [Eqn.~\eqref{eq:1}], a horizontal elongation and a vertical compression must take hold on this magnetic configuration. The configuration depicted on Fig.~\ref{fig:F0}(d) (``stripy AFM'') has alternating magnetic stripes with opposite Ising orientation \cite{cit:sivadasPRB2015}; there, a horizontal compression and a vertical elongation must take hold. The information in Fig.~\ref{fig:F0} holds regardless of which  structure is the magnetic ground state.

\begin{table*}
    \centering
    \caption{Crystalline layer group, total energy difference ($\Delta E$) with respect to zigzag AFM configuration, changes on diagonal ($u_d$) and vertical ($u_v$) distances between Cr atoms, and Voronoi charge transfers ($\Delta n$)---a positive value implies the atom gains electrons---for CrSiTe$_3$ MLs in four Ising configurations. The last two columns display the NM and our reference structure.\label{tab:parameters}}
    \begin{tabular}{c|c c c c c c}
        \hline \hline
Structure                                 & zigzag              & FM                 & N\'eel              & stripy               & NM                & reference           \\
                                          & AFM                 &                    & AFM                 & AFM                  & (spin-unpolarized)& (FM+N\'eel AFM)/2   \\
Layer group                               & $c2/m$              & $p\bar{3}1m$       & $p\bar{3}1m$        & $c2/m$               & $c2/m$            & $p\bar{3}1m$        \\
        \hline
        $\Delta E$ (meV/f.u.)             &  0.000              & 8.521              & 37.958              &  39.237              &  3.209$\times10^3$& ---                 \\
        \hline
        $a$ (\AA)                         &  6.991              & 6.984              &  6.941              &   6.931              &  6.888            & 6.962               \\
                                          & (+0.404\%)          & (+0.312\%)         &  ($-0.312$\%)       & ($-0.448$\%)         & ($-1.069$\%)      & (0.000\%)           \\
        $b$ (\AA)                         & 12.011              & ---                &  ---                &  12.095              & $a\sqrt{3}$       & ---                 \\
                                          & ($-0.394$\%)        & ---                &  ---                & ($+0.302$\%)         & ---               & ---                 \\
        $u_d$ (\AA)                       &$+0.024$             &$+0.013$            &$-0.013$             &$-0.021$              &$-0.043$           & 0.000               \\
        $u_v$ (\AA)                       &$-0.049$             &$+0.013$            &$-0.013$             &$+0.038$              &$-0.043$           & 0.000               \\
        \hline
        $\Delta n_{\text{Cr}}$ (e$^-$)    & $-0.242$            & $-0.247$           & $-0.238$            & $-0.242$             & $-0.305$          & ---                 \\
        $\Delta n_{\text{Si}}$ (e$^-$)    & $-0.305$            & $-0.301$           & $-0.306$            & $-0.302$             & $-0.308$          & ---                 \\
        $\Delta n_{\text{Te}}$ (e$^-$)    &  +0.182             &  +0.183	         &  +0.181	           &  +0.181              &  +0.204           & ---     	        \\
        \hline \hline
    \end{tabular}
\end{table*}

Table \ref{tab:parameters}  contains the total energies per formula unit (f.u.), layer groups, structural, and charge transfer information of four CrSiTe$_3$ ML magnetic Ising configurations; their total energy increases in the following progression: $E_{zigzag\text{}AFM}<E_{FM}<E_{N\acute{e}el\text{}AFM}<E_{stripy\text{}AFM}$~in consistency with Sivadas {\em et al.}~\cite{cit:sivadasPRB2015}; Wyckoff positions are reported in Ref.~\cite{SI}. The NM structure was created by turning spin polarization off \cite{cit:houmesNC2023} and it does not lead to a suitable reference. Indeed, similar to Ref.~\cite{cit:houmesNC2023}, the NM ML is compressed by 1.376\% with respect to the FM one, {\em and still compressed 0.759\% with respect to the N\'eel AFM one}: {\em in other words, the NM structure is inconsistent with the effect of magnetic forces on a reference lattice depicted on Figs.~\ref{fig:F0}(a,b), whereby the N\'eel AFM ML ought to compress.} {Furthermore}, there is a 3209 meV energy difference between the {NM} lattice and the zigzag AFM one, and the charge transfers between for Cr and Te atoms on the NM phase change on the leading decimal, as opposed to the second or third decimal listed for all the other four magnetic structures.

Houmes {\em et al.}~\cite{cit:houmesNC2023} contributed anisotropic changes in lattice parameters as a function of magnetic texturing for MPS$_3$ MLs, but { neglected to } explain the mechanics. As per the discussion in the previous paragraph, their NM structure does not lead to the proper elastic behavior of FM and N\'eel Ising textures. A better reference structure is the average lattice between FM and N\'eel configurations [away from which isotropic distortions are drawn in Figs.~\ref{fig:F0}(a) and \ref{fig:F0}(b)], for which the difference in $a$ is a meager $\pm 0.312$\% (written within parenthesis in Table \ref{tab:parameters}). Consistent with Figs.~\ref{fig:F0}(c,d), Table \ref{tab:parameters} indicates that the zigzag AFM lattice expands horizontally by 0.404\% and compresses by 0.394\% vertically. The stripy AFM lattice compresses by 0.448\% horizontally while expanding by 0.302\% in the vertical direction.

Figure~\ref{fig:F1}(a) shows the CrSiTe$_3$ ML on its ground state orthorhombic zigzag AFM configuration [primitive lattice vectors are: $\mathbf{a}=a(1,0,0)$ and $\mathbf{b}=b(0,1,0)$]. Fig.~\ref{fig:F1}(b) depicts the ML on its FM configuration [$\mathbf{a}=a\left(-\frac{1}{2},-\frac{\sqrt{3}}{2},0\right)$ and $\mathbf{b}=a(1,0,0)$]. Additional details of atomic positions are provided in Ref.~\cite{SI}. In agreement with a Biot-Savart picture, a {\em bona fide} lattice to calculate magnetoelastic distortions is established now.

\begin{figure}[tb]
    \centering
    \includegraphics[width=0.45\textwidth]{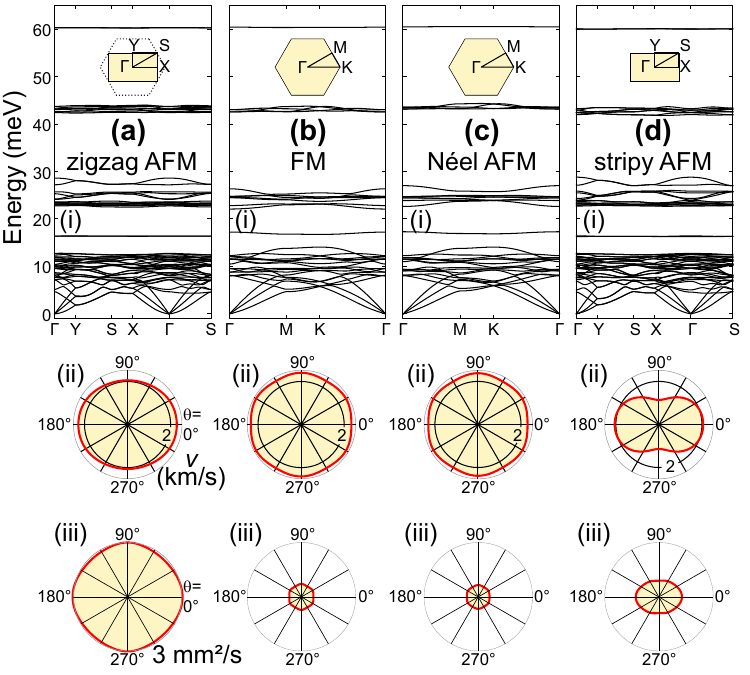}
    \caption{{(i)} Phonon dispersions, {(ii)} transverse group velocities, and {(iii)} quadratic fitting parameters as a function of angle $\theta$ (0$^\circ$ corresponds to the $x$-direction).\label{fig:phonons}}
\end{figure}

To explain the anisotropy in fundamental frequencies experimentally observed on similar {(finite-size)} magnets \cite{cit:houmesNC2023}, the vibrational spectrum of the four magnetic phases is displayed on Figs.~\ref{fig:phonons}(a-d)(i). A lack of imaginary vibrational frequencies implies that the four structures are stable. Figs.~\ref{fig:phonons}(a-d)(ii) display the speed of sound: the two-fold symmetric shape observed on the zigzag AFM and stripy AFM MLs underpins anisotropic fundamental frequencies \cite{cit:houmesNC2023}. The quadratic fitting parameter on Figs.~\ref{fig:phonons}(a-d)(iii) is anisotropic for the zigzag AFM and stripy AFM configurations as well.

The diagonalized stiffness between nearest Cr ions on the FM (N\'eel AFM) ML are 0.341 (0.215) eV/\AA$^{2}$ for the longitudinal ($K_L$) mode,  $-$0.053 ($-$0.045) eV/\AA$^{2}$ for the in-plane transverse ($K_{T,i}$) mode, and $-$0.167 ($-$0.186) eV/\AA$^{2}$ for the out-of-plane transverse ($K_{T,o}$) mode. Changes from an ``average stiffness'' are $\pm 22.7$\%, $\pm 8.5$\%, and $\mp 5.3$\% for $K_L$, $K_{T,i}$, and $K_{T,o}$, respectively. More details are available in Ref.~\cite{SI}. The changes in stiffness between the FM and N\'eel MLs are reliable indicators of the magnetization-dependent {\em hardening} or {\em softening} of the lattice.

Having a valid reference lattice permits calculating the elastic energy required to turn into the four magnetic structures away from the reference structure. Rectangular cells with 20 atoms were used for the four magnetic phases for direct comparison. Atomic displacements $\mathbf{u}_i$ for each atom $i$ within a unit cell on a given magnetic phase onto the {non-magnetic} average structure were calculated as row vectors and $\textbf{u}=(\mathbf{u}_1,\mathbf{u}_2,...,\mathbf{u}_{20})^T$. In this calculation, the origin coincides with the center of mass of {\em both} (reference and a given magnetic) structures, {making $\mathbf{u}$ and the lattice vectors of a given magnetic phase interrelated}. The elastic energy is a quadratic form:
\begin{equation}\label{eq:3}
	E_{el}=\frac{1}{2}\textbf{u}^T\mathcal{K}\textbf{u},
\end{equation}
where $\mathcal{K}$ is the Hessian used to obtain each of the phonon dispersions in Fig.~\ref{fig:phonons}. Since the motion from the reference to a magnetic structure is {non-dispersive} (no sound is involved in such distortion), the elastic energy was evaluated at $\Gamma$, even though elastic interactions up to multiple neighbors were considered, and ``folded'' onto the unit cell using Bloch's theorem. As seen in Table~\ref{tab:elasticity}, the elastic energy is no larger than 3.690 meV/f.u. In contrast, the elastic energy is overestimated to be over 109 meV/f.u.~when using the NM structure as reference (last row on Table \ref{tab:elasticity}). Magnetic parameters $\mathbf{J}_{ij}$, $\mathbf{D}_{ij}$ and $\mathbf{A}_{ij}$ [Eqn.~\eqref{eq:heisenberg}] were computed for the four magnetic phases;  $\mathbf{J}(r)$ remained significant even at distances of 12~\AA{}, and the magnetic energy $E_m$ was estimated using Ising spins and the magnetic exchange provided in Ref.~\cite{SI}. The sum of elastic and magnetic energies for the stripy AFM ML happens to be 38.290 meV larger than that of the zigzag AFM ML, consistent with Table \ref{tab:parameters}.


\begin{table}[tb]
    \centering
    \caption{Elastic deformation energies away from the reference structure. Also listed are deformation energies using the NM structure and the Ising magnetic energies (meV/f.u.).\label{tab:elasticity}}
    \begin{tabular}{c|cccc}
        \hline \hline
                                & zigzag AFM    & FM        & N\'eel AFM    & stripy AFM \\
        \hline
        \hline
        $E_{el}$           &   3.690       &   3.446   &   3.447       &   1.807 \\
        $E_m$              & -25.260       &   4.590   &   0.810       &  14.910 \\
        $E_m+E_{el}$       & -21.570       &   8.036   &   4.257       &  16.717 \\ 
\hline
        $E_{el,NM}$        & 109.572       & 120.007   & 124.990       & 138.050 \\
        \hline \hline
    \end{tabular}
\end{table}

\begin{table}[tb]
    \centering
    \caption{Elastic parameters (eV/\AA) and elastic energies (meV/f.u.).}
    \begin{tabular}{c|cccc}
        \hline \hline
                                & zigzag AFM    & FM        & N\'eel AFM    & stripy AFM \\
        \hline
        \hline
        $C_{11}$ (eV/\AA{}$^2$) &   3.153       &   3.211   &   2.928       &   3.063 \\
        $C_{22}$ (eV/\AA{}$^2$) &   2.753       & $C_{11}$  & $C_{11}$      &   3.084 \\
        $C_{12}$ (eV/\AA{}$^2$) &   0.647       &   0.689   &   0.584       &   0.569 \\
        $C_{66}$ (eV/\AA{}$^2$) &   1.191       &   1.261   &   1.172       &   1.207 \\
        $U_{el}$    (meV)       &   3.162       &   3.188   &   2.871       &   3.073 \\
        $E_m+U_{el}$ (meV)      & -22.098       &   7.778   &   3.681       &  17.983 \\ 
        \hline \hline
    \end{tabular}
    \label{tab:elasticity2}
\end{table}

Elastic moduli were calculated for each magnetic phase using rectangular cells. Tensile strain $\epsilon_i=+0.5$\% ($i=x,y$) was applied along $a$ and $b$ to calculate $C_{11}$, $C_{22}$, and $C_{12}$; a shear strain of 0.9\% was used to calculate $C_{66}$. For $p\bar{3}1m$ and $c2/m$ structures, $C_{16}=C_{26}=0$. Additionally, $C_{11}=C_{22}$ and $C_{66}=(C_{11}-C_{12})/2$ for FM and N\'eel AFM MLs, requiring only a single strained-cell calculation to decouple the elastic moduli. The results are placed in Table \ref{tab:elasticity2}. As expected, the FM lattice is shown to be more rigid than AFM lattices. More details, as well as comparable values calculated from group velocities (as shown in Fig. \ref{fig:phonons}) can be found in Ref. \cite{SI}. The elastic energy can be similarly estimated by modifying the elastic energy density formula:
\begin{equation} 
    u_{el} = \frac{1}{2}C_{11}\epsilon_{x}^2 + \frac{1}{2}C_{22}\epsilon_{y}^2+C_{12}\epsilon_x\epsilon_y,
\end{equation}
by multiplying by the unperturbed cells' area $A=ab$, such that the elastic energy is equal to $U_{el}=u_{el}A$. The results, listed in Table \ref{tab:elasticity2}, are similar to those obtained using Eqn.~\eqref{eq:3}.

\begin{figure}[tb]
    \centering
    \includegraphics[width=0.45\textwidth]{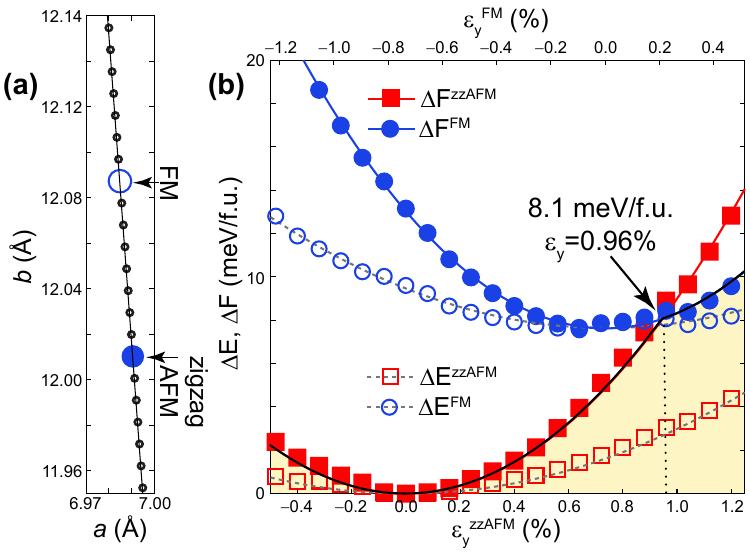}
    \caption{(a) Lattice constants $a$ and $b$ joining the zigzag AFM and the FM phases. (b) Total energy and free energy paths: the free energies cross under stress, and indicate a magnetic quantum phase transition at 0.96\% vertical strain.\label{fig:pt}}
\end{figure}

To end this work, we predict the amount of uniaxial strain needed to switch the zigzag AFM ML onto the FM one \cite{free-energies}. The strain needed to link the local minima of those two magnetic phases is shown on Fig.~\ref{fig:pt}(a), and the free energy is defined as:
\begin{equation}
    F=E^{(i)}-V^{(i)}\sum_{j=x,y}\sigma_{jj}^{(i)}\epsilon_j,
\end{equation}
where $E^{(i)}$ is the total energy for the zigzag AFM or for the FM phase [(i): zzAFM, or FM], $V$ is the volume of the computational cell, and $\sigma_{jj}$ is the remanent normal stresses on the lattice. The total and free energies, relative to zigzag AFM ground state, are plotted in Fig.~\ref{fig:pt}(b). An 8.1 meV/f.u. barrier is crossed with a strain of 0.96\%, and an AFM to FM quantum phase transition is thus achieved under tensile strain.

In conclusion, we motivated and established a meaningful reference lattice structure from which to calculate magnetoelastic energies. We contributed space groups for four magnetic phases, provided sound speeds to be tested experimentally and phonon dispersion relations for four magnetic phases, calculated changes in stiffness due to magnetism, estimated stiffness parameters and elastic moduli, reported  elastic and magnetic energies, all to be found of the order of a few meV, and discarded a NM (spin-unpolarized) lattice as a meaningful reference for magnetoelasticity. Furthermore, we estimate an AFM to FM quantum phase transition under 0.96\% tensile strain along the long lattice vector. The original insight of a lattice of solenoids leads onto a transparent description of magnetoelasticity as a perturbation from a sensible reference lattice. This comprehensive study thus provides insightful vistas into magnetoelastic couplings in 2D Ising magnets.

\acknowledgments{We acknowledge funding from the DOE (S.B.L.: contract DE-SC0022120; J.M.D.: contract DE‐SC0014664). A.G.-F.~and J.F.~were funded by project PID2022-137078NB-I00 (MCIN/AEI/10.13039/501100011033/FEDER, EU). DFT calculations were performed at the Pinnacle Supercomputer, funded by NSF under Award No. OAC-2346752, and at NERSC (contract number DE-AC02-05CH11231; NERSC award BES-ERCAP0031261). We thank Angiolo Huam\'an for useful discussions.


\end{document}